\documentclass[slac_one]{revtex4}
\usepackage{graphicx}
\usepackage{fancyhdr}
\def\itbf#1{\mbox{\boldmath $#1$}}
\def\sitbf#1{\mbox{\scriptsize\boldmath $#1$}}

\def\psip#1{\psi_{\mathbf{#1}}}
\def\chip#1{\chi_{\mathbf{#1}}}

\def\lsim{\mathrel{\raise.3ex\hbox{$<$\kern-.75em\lower1ex\hbox{$\sim$}}}}
\def\gsim{\mathrel{\raise.3ex\hbox{$>$\kern-.75em\lower1ex\hbox{$\sim$}}}}

\newcommand{\bbmp}{\mbox{\scriptsize\boldmath $p$}}

\catcode`\@=11
\def\slash{\mathpalette\make@slash}
\def\make@slash#1#2{\setbox\z@\hbox{$#1#2$}%
  \hbox to 0pt{\hss$#1/$\hss\kern-\wd0}\box0}
\catcode`\@=12 

\begin{document}

\title{{\small{2005 International Linear Collider Workshop - Stanford,
U.S.A.}}\\ 
\vspace{12pt}
Effects of the $t\bar t$ threshold in $e^+e^-\rightarrow t\bar t H$}

\author{C. Farrell, A.H. Hoang}
\affiliation{Max-Planck-Institut for Physics,
Munich, Germany}

\begin{abstract}
In the region where the Higgs energy is large 
the process $e^+e^-\to t\bar t H$ is governed by nonrelativistic QCD
dynamics and one has to employ effective theory methods to make first
principles QCD predictions. In this talk we use the effective theory
vNRQCD to compute the Higgs energy distribution at 
next-to-leading logarithmic approximation. It is shown that the corrections are
particularly important for smaller c.m.\,energies.
\end{abstract}

\maketitle

\thispagestyle{fancy}

\section{Introduction}
\label{sectionintroduction}

It is one of the major tasks of future collider experiments to unravel
details of the mechanism of electroweak symmetry breaking (EWSB). In the
Standard Model (SM) EWSB is achieved   
by the Higgs mechanism. The particle masses are generated by the
Higgs field vacuum expectation value $V=(\sqrt{2}\,G_F)^{1/2}\approx
246$~GeV arising through the Higgs self interactions. The
mechanism also predicts that Higgs bosons can be produced in
collider experiments. While a Higgs boson with a mass smaller than $1$~TeV
can be found at the LHC, precise and model-independent measurements of its
quantum numbers and couplings can be gained from the $e^+e^-$ Linear
Collider. A crucial prediction of the Higgs mechanism is that the Higgs Yukawa coupling to
quarks $\lambda_q$ is related to the quark masses by   $m_q=\lambda_q
V$. At the $e^+e^-$ Linear Collider the top quark Yukawa coupling
can be measured from top quark pair production associated with a
Higgs boson, $e^+e^-\to t\bar t H$, since the process is dominated by
the amplitudes describing Higgs radiation off the $t\bar t$ pair. This
process is particularly suited for a light Higgs boson since the cross
section can then reach the $1$-$2$~fb level. Assuming an experimental
precision at the percent level, QCD and electroweak radiative
corrections need to be accounted for in the theoretical predictions.  
In the approximation that top quarks and the Higgs are stable
particles the Born cross section was already determined some time ago in
Refs.~\cite{Borneetth}. For the ${\cal O}(\alpha_s)$ QCD one-loop
corrections a number of references in various approximations exist
\cite{Dawson1,Dawson2,Dittmaier1}. On the other hand, the 
full set of one-loop electroweak corrections was obtained in
Refs.~\cite{Belanger1,Denner1} and also in Refs.~\cite{You1}. In
Ref.~\cite{Denner1} a detailed analysis of various differential 
distributions of the cross section $\sigma(e^+e^-\to t\bar t H)$ can
be found.

%
%
\begin{figure}[t] 
  \begin{center}
    \includegraphics[width=6cm]{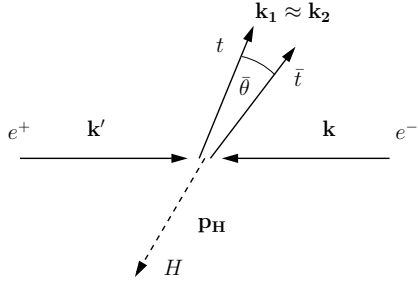}
    \vskip  0.0cm
    \caption{
      Typical constellation of momenta for the process $e^+e^-\to t\bar t H$ 
      in the large Higgs energy endpoint region. 
      \label{fig1} }
  \end{center}
\end{figure}
A particularly
interesting kinematical phase space region is where the energy of the Higgs
boson is large and close to its kinematic endpoint. The $t\bar t$
pair then becomes collinear and flies opposite
to the Higgs direction  to balance the large Higgs momentum, see
Fig.~\ref{fig1}. For large $E_H$, on the other hand, the $t\bar t$
invariant mass $Q^2$ approaches $4m_t^2$, $E_H \, = \,
\frac{1}{2\sqrt{s}}\,\left(s+m_H^2-Q^2\right)$, and the top quark pair is
nonrelativistic in its own center-of-mass (c.m.) system. Because the
Higgs is very narrow for a mass below the $W^+W^-$ threshold, strong
interactions between the 
$t\bar t$ pair and the hadronic Higgs final state can be neglected.  
Thus, close to the Higgs  energy endpoint the $t\bar t$ QCD dynamics
is exclusively governed by the nonrelativistic physics known from the
process $e^+e^-\to t\bar t$ in the $t\bar t$ threshold region at
$\sqrt{s}\approx 2m_t$.  
In this regime the so-called Coulomb singularities $\propto (\alpha_s/v)^n$,
with $v=(1-4m_t^2/Q^2)^{1/2}$ being the top quark relative velocity in the
$t\bar t$ c.m.\,frame, arise and require predictions using an
expansion in $\alpha_s$ and $v$ rather than just a perturbative computation
in the number of loops.  

This singularity structure is
most easily visible in the Higgs energy distribution, $d\sigma(e^+e^-\to t\bar
t H)/d E_H$. While the Born distribution approaches zero for 
$E_H\to E_H^{\rm max}$, $d\sigma/d E_H\sim v$~\cite{Borneetth},
the ${\cal O}(\alpha_s)$ fixed-order perturbative corrections are proportional 
to $\alpha_s$ at the endpoint~\cite{Dawson2,Dittmaier1} and the ${\cal O}(\alpha_s^2)$
corrections even diverge like 
$\alpha_s^2/v$. The problem might be avoided by imposing a cut on
$E_H$ or $Q^2$, but this is unnecessary because there exists an 
elaborate technology being used for the threshold region in
the process $e^+e^-\to t\bar t$~\cite{TTBARreview} that allows for
systematic QCD predictions with renormalization group (RG)
improvement. Imposing a cut would be also disadvantageous as the
nonrelativistic portion of the $t\bar t H$ phase space increases  
for smaller c.m.\,energies which are relevant for a measurement in the first
phase of the ILC program. Because the SM top width is quite large,
$\Gamma_t\approx 1.5$~GeV, the corresponding QCD effective theory
computations can be carried out with perturbative methods for all
Higgs energies in the endpoint region.   

In this talk we present the Higgs energy distribution $d\sigma/d E_H$
in the large Higgs energy endpoint region at NLL order in the nonrelativistic
expansion using the framework of ``velocity'' NRQCD
(vNRQCD).  For    
details on the conceptual aspects, concerning powercounting, the operator
structure of the effective theory action, and renormalization we refer to
Refs.~\cite{LMR,amis,amis2,HoangStewartultra,hmst}. 
For a more detailed discussion of the computations for this work see
Ref.~\cite{farrellhoang1}.

\section{Review of Effective Theory Ingredients} 
\label{sectionEFT}

The effective theory vNRQCD provides a systematic RG-improved
description of  dynamics of nonrelativistic $t\bar t$ pairs. The
system is characterized, for any energy in the threshold region, by the hierarchy
\begin{eqnarray}
m_t \gg m_t v~\mbox{(three-momentum, ``soft'' scale)} 
\gg m_t v^2~\mbox{(kinetic energy, ``ultrasoft'' scale)}
\gg \Lambda_{\rm QCD}
\,.
\end{eqnarray}
The particle-antiparticle propagation is described by the terms in the
effective theory Lagrangian which are bilinear in the top quark and antitop quark
fields,
\begin{eqnarray} \label{Lke}
 {\mathcal L}(x) &=& \sum_{\bbmp}
   \psip{\bbmp}^\dagger(x)   \biggl\{ i \partial^0 - {\itbf{p}^2 \over 2 m_t}   
   + \frac{i}{2} \Gamma_t 
   - \delta m_t \biggr\} \psip{\bbmp}(x) 
+ (\psip{\bbmp}(x) \to\chip{\bbmp}(x))
\,,
\end{eqnarray}
where the fields $\psip{\bbmp}$ and $\chip{\bbmp}$ destroy top and
antitop quarks with soft three-momentum ${\itbf{p}}$ in the $t\bar t$
c.m.\,\,frame and $\Gamma_t$ is the 
on-shell top quark decay width. The term $\delta m_t$ is a residual mass term
specific to the top quark mass definition that is being used; for our
analysis we employ the 1S mass scheme~\cite{Hoangupsilon,HoangTeubnerdist}. 

Up to NLL order the top-antitop quark pair interacts only through the effective
Coulomb potential~\cite{FischlerBilloire},
\begin{eqnarray}
 \tilde V_c({\itbf{p}},{\itbf{q}}) 
 & = &
- \frac{4\pi C_F \alpha_s(m_t \nu)}{{\itbf{k}}^2}\, \left\{\,1 + 
 \frac{\alpha_s(m_t\nu)}{4\pi}\,\left[\,
 -\beta_0\,\ln\Big(\frac{{\itbf{k}}^2}{m_t^2\nu^2}\Big) + a_1
 \,\right]\,\right\} 
\,,
 \label{VCoulomb}
\end{eqnarray}
where ${\itbf{k}}={\itbf{p}}-{\itbf{q}}$ is the momentum transfer and
$\beta_0=11/3 C_A-4/3 T n_f$ is the one-loop QCD beta function, 
$a_1= 31/9 C_A - 20/9 T n_f$ the coefficient of the one-loop correction to the
effective Coulomb potential, and $C_A=3, C_F=4/3, T=1/2$ are SU(3) group
theoretical factors. For the number of light flavors we take $n_f=5$. The
parameter $\nu$ is the vNRQCD renormalization scaling parameter used
to describe the correlated running of soft and ultrasoft effects in
the effective theory. Here, $\nu=1$ corresponds to the hard
scale at which the effective theory is matched to the full theory, and
$\nu=v_0$, $v_0$ being of the order of the typical $t\bar t$ relative
velocity, is the scale where the matrix elements are computed. The evolution
of the Wilson coefficients from the matching scale down to the low-energy
scale sums logarithms of the velocity to all orders and
is governed by the velocity renormalization group equations~\cite{LMR}.

Top-antitop quark production in the nonrelativistic regime in the LL and NLL
approximation in a ${}^3S_1$ spin triplet or a ${}^1S_0$ spin singlet state
is described by the currents 
\begin{equation}
  J^j_{1,{\sitbf{p}}}  = 
    \psi_{{\sitbf{p}}}^\dagger\, \sigma_j (i\sigma_2) \chi_{-{\sitbf{p}}}^*
   \,,\qquad\qquad
  J_{0,{\sitbf{p}}}  =   \psi_{{\sitbf{p}}}^\dagger\,  (i\sigma_2)
    \chi_{-{\sitbf{p}}}^*
\,, 
\label{J1J0}
\end{equation}
where $c_{1,j}(\nu)$ and $c_0(\nu)$ are the corresponding Wilson coefficients.
The currents do not run at LL order but UV-divergences in effective theory
two-loop vertex diagrams~\cite{LMR} lead to
non-trivial anomalous dimensions at NLL order which result in a
scaling of the  Wilson coefficients~\cite{HoangStewartultra,Pineda1}
\begin{equation}
c_{1,j}(\nu)  =  c_{1,j}(1)\,\exp\left(f(\nu,{\bf S^2}=2) \right) 
   \,,\qquad\qquad
c_{0}(\nu)  =  c_{0}(1)\,\exp\left(f(\nu,{\bf S^2}=0) \right)\,,
\label{currentWilson}
\end{equation}
where ${\bf S^2}$ is the square of the total $t\bar t$ spin. We
have adopted the 
convention that the matching conditions at $\nu=1$ only account for QCD
effects, so at LL order we have $c_{1}(1)=c_{0}(1)=1$. The NLL order QCD
matching conditions relevant for $e^+e^-\to t\bar t H$  in the large Higgs
energy endpoint region are discussed in Sec.~\ref{sectionmatching}.  

Through the optical theorem the $t\bar t$ production rate for a $t\bar t$
invariant mass $Q^2\approx 4m_t^2$ involves the imaginary part of the
time-ordered product of the production and annihilation currents defined in
Eqs.~(\ref{J1J0}),
\begin{eqnarray}
 {\cal A}_1^{lk}(Q^2,m_t,\nu) & = & i\,
 \sum\limits_{\mbox{\scriptsize\boldmath $p$},\mbox{\scriptsize\boldmath $p'$}}
 \int\! d^4x\: e^{-i \hat{q} \cdot x}\:
 \Big\langle\,0\,\Big|\, 
    T\, J^{l\dagger}_{1,{\sitbf{p}^\prime}}(0) J^k_{1,{\sitbf{p}}}(x)
 \Big|\,0\,\Big\rangle 
\,=\,2 N_c \delta^{lk}\, G^c(a,v,m_t,\nu)\,,\quad
\label{A1def}
\\[2mm]
 {\cal A}_0(Q^2,m_t,\nu) & = & i\,
 \sum\limits_{\mbox{\scriptsize\boldmath $p$},\mbox{\scriptsize\boldmath $p'$}}
 \int\! d^4x\: e^{-i \hat{q} \cdot x}\:
 \Big\langle\,0\,\Big|\, 
    T\, J^\dagger_{0,{\sitbf{p}^\prime}}(0) J_{0,{\sitbf{p}}}(x)
 \Big|\,0\,\Big\rangle 
\, = \, N_c\, G^c(a,v,m_t,\nu)\,,
\label{A2def}
\end{eqnarray}
where $ v  = ((\sqrt{Q^2}-2 m_t-2\delta m_t+i\Gamma_t)/m_t)^{\frac{1}{2}}$
is the c.m.\,\,top quark (effective) relative velocity and
$\hat{q}\equiv(\sqrt{Q^2}-2m_t,0)$. The term $G^c$ is the zero-distance S-wave
Coulomb Green function of the nonrelativistic Schr\"odinger equation with the 
potential in Eq.~(\ref{VCoulomb}). 
To compute the Green function we
use the numerical techniques and codes of the TOPPIC program developed in
Ref.~\cite{Jezabek1} and  
determine an exact solution of the full NLL Schr\"odinger equation following
the approach of Refs.~\cite{hmst}.  

\section{Comments on the Computation} 
\label{sectionmatching}

The LL vNRQCD result for the Higgs energy distribution, including the QCD
effects coming from the Coulomb potential in Eq.~(\ref{VCoulomb}),
the finite top quark lifetime, and the Wilson coefficients of the
currents in Eqs.~(\ref{currentWilson}), is given by
\begin{eqnarray}
\frac{d\sigma}{d E_H}(E_H\approx E_H^{\rm max}) & = &
\frac{8\,N_c\,\left[(1+x_H-4x_t)^2-4x_H\right]^{1/2}}{s^{3/2}\,m_t^2}\,
\left(\,c^2_0(\nu)\, F^Z_0 + c_1^2(\nu)\,F^{\gamma,Z}_1\,\right)\,
\,\mbox{Im}\left[\, G^c(a,v,m_t,\nu)\,\right]\,
\,.
\label{dsdEHEFT}
\end{eqnarray}
For details on the explicit calculation and the definition of the formfactors $F_i$
see~\cite{farrellhoang1}. We note that the NLL (${\cal O}(\alpha_s)$)
matching conditions for the three triplet Wilson coefficients
$c_{1,j}$ depend on 
the $t\bar t$ spin configuration (i.e.\,\,on $j$) since the kinematic
situation for $t\bar t H$ production in the large Higgs energy endpoint is not
invariant under separate rotations of the spin quantization axis. However, for
our purposes it is sufficient to define a triplet Wilson coefficient that is
averaged over the three spin configurations. Using such an averaged triplet
Wilson coefficient the Higgs energy spectrum at NLL order can also be cast in
the simple form of Eq.~(\ref{dsdEHEFT}). 

At NLL order, we need
to account for the  ${\cal O}(\alpha_s^2)$ contributions to the Coulomb
potential in Eq.~(\ref{VCoulomb}), the NLL running of the coefficients $c_1$
and $c_0$, and their ${\cal O}(\alpha_s)$ matching conditions at $\nu=1$. The
latter hard QCD corrections are process specific and cannot be inferred from
results obtained in earlier computations for the $t\bar t$ threshold in
$e^+e^-\to t\bar t$.    
%
%
\begin{figure}[t] 
  \begin{center}
    \includegraphics[width=10cm]{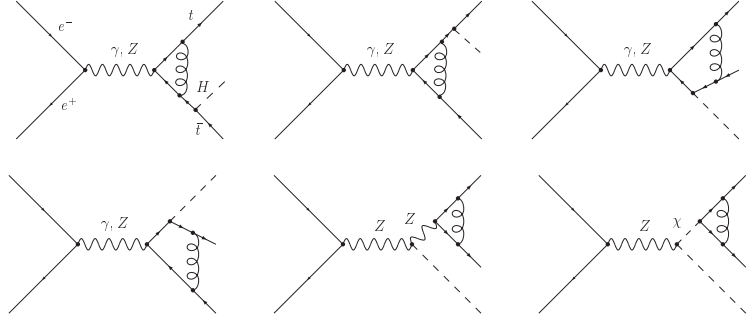}
    \caption{
      Diagrams describing virtual one-loop QCD corrections in the Standard 
      Model to compute the NLL matching conditions for the operators  
      $J_{1,{\sitbf{p}}}$ and $J_{0,{\sitbf{p}}}$ that describe $t\bar t$ 
      production for invariant masses $Q^2\approx 4m_t^2$. Self energy 
      diagrams are implied.
 \label{fig2} }
\end{center}
\end{figure}
We have extracted the matching conditions 
from the codes for the Standard Model amplitude for 
$e^+e^-\to t\bar t H$ provided in Ref.~\cite{Denner1}. With the ansatz 
\begin{equation}
c_{0,1}(\nu=1,\sqrt{s},m_t,m_H) = 1 + \frac{C_F\alpha_s(m_t)}{\pi}\,
\delta c_{0,1}(\sqrt{s},m_t,m_H)
\label{matchcond}
\end{equation}
we have determined the ${\cal O}(\alpha_s)$ matching conditions
numerically by matching the  ${\cal  O}(\alpha_s)$ vNRQCD prediction 
at $\mu=m_t$ ($\nu=1$) to the full theory results close to the large
Higgs energy endpoint. The relative uncertainties for this numerical
procedure are below 1\% for $\delta c_{0,1}$.

\section{Numerical Analysis} 
\label{sectionanalysis}

%
%
\begin{figure}[t] 
  \begin{center}
    \includegraphics[width=8cm]{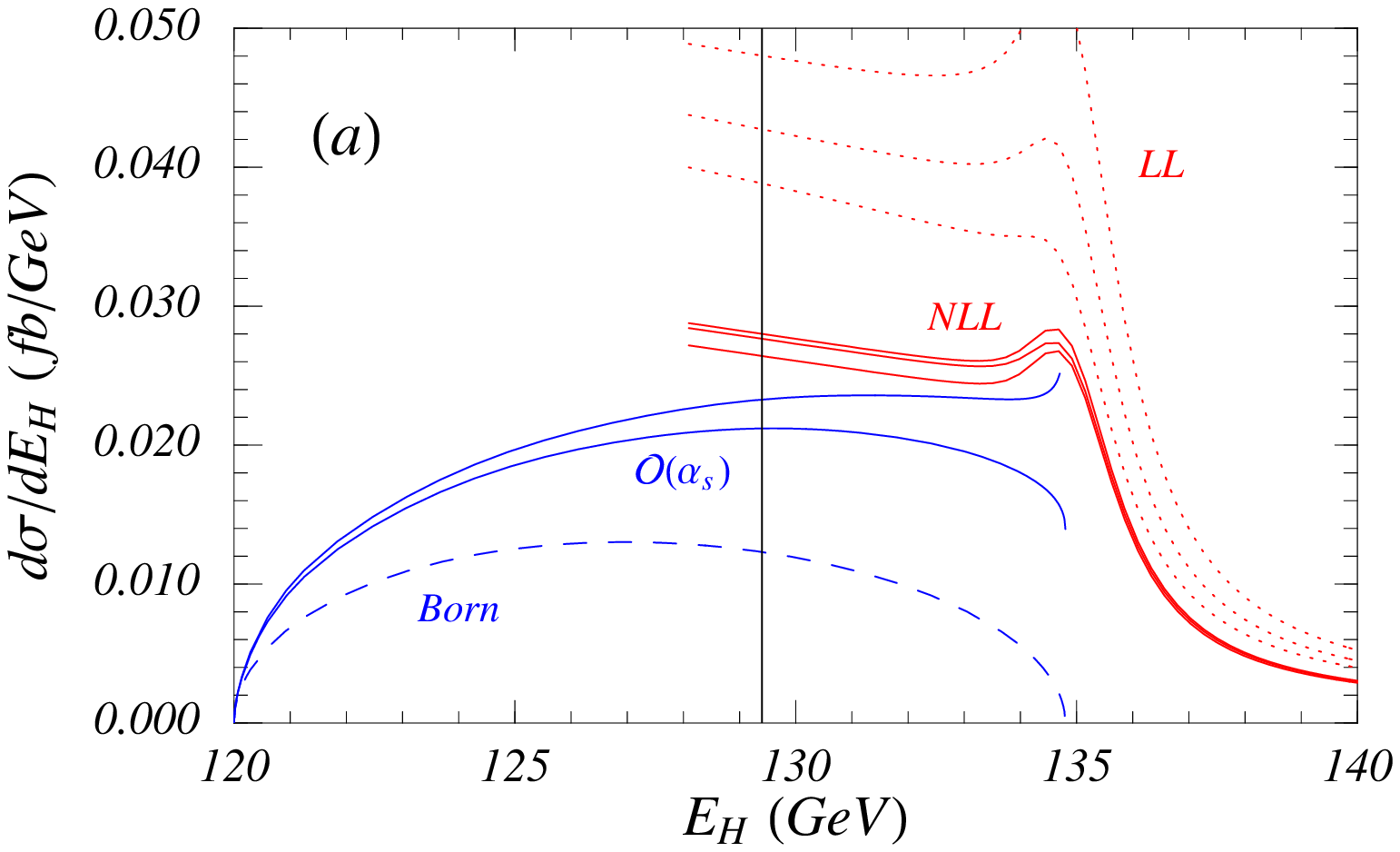}
    \includegraphics[width=8cm]{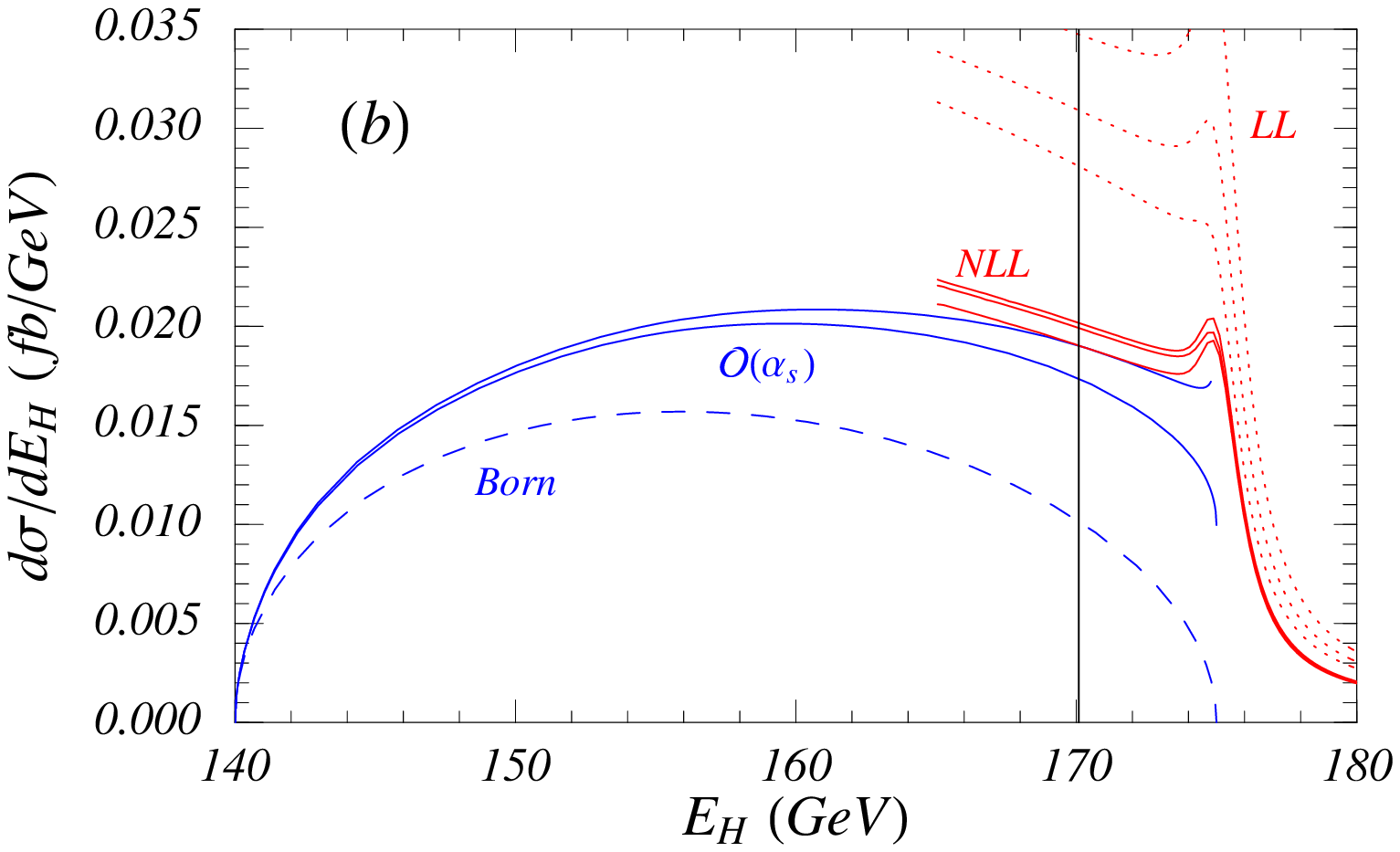}
\vskip -0.3cm
    \caption{
      Higgs energy spectrum at LL (dotted lines) and 
      NLL (solid lines) order in the nonrelativistic expansion
      for the vNRQCD renormalization parameters $\nu=0.1,0.2,0.4$
      and at the Born level and at ${\cal O}(\alpha_s)$ for
      $\mu=\sqrt{s}, \sqrt s |v|$ for the parameters
      (a) $\sqrt{s}=500$~GeV, $m_H=120$~GeV and
      (b) $\sqrt{s}=550$~GeV, $m_H=140$~GeV.
      The top 1S mass has been set to $m_t^{\rm 1S}=180$~GeV and the
      other parameters are $\Gamma_t=1.55~\mbox{GeV}$, 
      $M_Z=91.1876$~GeV, $M_W=80.423$~GeV, $c_w=M_W/M_Z$ and
      $\alpha^{-1}=137.034$. 
      \label{fig3} }
  \end{center}
\end{figure}

\begin{table}
\begin{center}
\begin{tabular}{|c|c||c|c||l||c|c|c|}
\hline
$\sqrt{s}$ [GeV] & 
$m_H$ [GeV] & 
$\mbox{}\;\sigma(\mbox{Born})$ [fb] & 
$\mbox{}\;\sigma(\alpha_s)$ [fb] &
$\mbox{}\;\sigma(\mbox{NLL})$ [fb] & 
$\mbox{}\quad\frac{\sigma(\mbox{\tiny NLL})}{\sigma(\mbox{\tiny Born})}\quad\mbox{}$ &
$\mbox{}\quad\frac{\sigma(\mbox{\tiny
    NLL})}{\sigma(\alpha_s)}\quad\mbox{}$ &
$\frac{\sigma(\mbox{\tiny NLL})_{|\beta|<0.2}}{\sigma(\alpha_s)_{\beta<0.2}}$
\\ \hline\hline
$500$ & $120$ & $ 0.151$ & $ 0.263$ & $\quad 0.357(20)$ & $ 2.362$ & $
1.359$ & $ 1.78$\\\hline
$550$ & $120$ & $ 0.984$ & $ 1.251$ & $\quad 1.342(37)$ & $ 1.364$ & 
$ 1.073$  & $ 1.66$ \\\hline
$550$ & $160$ & $ 0.134$ & $ 0.207$ & $\quad 0.254(12)$ & $ 1.902$ & 
$ 1.226$ & $ 1.74$ \\\hline
$600$ & $120$ & $ 1.691$ & $ 1.939$ & $\quad 2.005(30)$ & $ 1.185$ & 
$ 1.034$ & $ 1.66$ \\\hline
$600$ & $160$ & $ 0.565$ & $ 0.700$ & $\quad 0.745(18)$ & $ 1.319$ & 
$ 1.065$ & $ 1.68$ \\\hline
$700$ & $120$ & $ 2.348$ & $ 2.454$ & $\quad 2.485(13)$ & $ 1.058$ & 
$ 1.012$ & $ 1.68$ \\\hline
$700$ & $160$ & $ 1.210$ & $ 1.303$ & $\quad 1.328(11)$ & $ 1.098$ & 
$ 1.020$ & $ 1.69$ \\\hline
\hline
\end{tabular}
\end{center}
{\tighten \caption{
Collection of cross sections and K factors for various c.m.\,energies
and Higgs masses and top quark mass $m_t^{\rm 1S}=180$~GeV. Here,
$\sigma(\mbox{Born})$  
refers to the Born cross section and  $\sigma(\alpha_s)$ to the 
${\cal O}(\alpha_s)$ cross section in fixed-order perturbation theory
using $\mu=\sqrt{s}$ as the renormalization scale, the choice employed
in the analysis of Ref.\,\cite{Denner1}. The term $\sigma(\mbox{NLL})$
refers to the sum of the ${\cal O}(\alpha_s)$ fixed-order cross
section for $v>0.2$ using $\mu=\sqrt{s}v$ and the NLL nonrelativistic
cross section for $|v|<0.2$ with the renormalization parameter
$\nu=0.2$. (See Figs.\,\ref{fig3} where the vertical lines indicate
the Higgs energy where $v=0.2$.)}
\label{tab3} }
\end{table}

In Figs.\,\ref{fig3} the predictions of the Higgs energy spectrum
in the full kinematic range are displayed for two cases. In the large
Higgs energy endpoint region we have shown the LL (dashed lines) and NLL (solid
lines) results in the nonrelativistic expansion. (See the figure
captions for details on choice of  
parameters.) At LL order the upper (lower) curve corresponds to
$\nu=0.1$ ($0.4$), while at NLL order the upper (lower) curve
corresponds to $\nu=0.2$ ($0.1$).  
The curves show the typical behavior of the prediction of the
nonrelativistic expansion for any choice of parameters. While 
the LL predictions have a quite large renormalization parameter
dependence at the level of several tens of percent, the NLL results
are stable. Here, the variation due to change of the renormalization
parameter is around
5\%. The stabilization with respect to renormalization parameter
variations at NLL order arises mainly from the inclusion
of the ${\cal O}(\alpha_s)$ QCD corrections to the Coulomb potential,
Eq.\,(\ref{VCoulomb}).  
Moreover, the NLL curves lie considerably lower than the LL ones. This
behavior is well known from the predictions for $e^+e^-\to t\bar t$ at
threshold~\cite{TTBARreview,hmst} and originates from the structure of the 
large negative  ${\cal O}(\alpha_s)$ QCD corrections to the Coulomb potential,
Eq.\,(\ref{VCoulomb}), and from the sizeable negative
QCD corrections to the matching conditions in
Eq.\,(\ref{matchcond}).

In principle this behavior is a point of concern because it could
indicate that the renormalization parameter variation might be an
inadequate method to estimate theoretical uncertainties. Fortunately,
the top quark mass is sufficiently large such that the regions where
the conventional fixed-order expansion (in powers of the strong
coupling) and where the nonrelativistic expansion (described by the
effective theory) can be applied are expected to overlap. To
demonstrate this issue we have also displayed in Figs.\,\ref{fig3} the
fixed-order (i.e. without summation of Coulomb singularities)
predictions at the Born (dashed lines)~\cite{Dawson1} and 
the ${\cal O}(\alpha_s)$ level (solid lines)~\cite{Denner1}. 
The two ${\cal O}(\alpha_s)$ curves correspond to the renormalization
scales $\mu=\sqrt{s}$ (lower curves) and $\mu=|\sqrt{s} v|$ (upper curves),
where $v$ is the $t\bar t$ relative velocity defined below
Eq.\,(\ref{A2def}). The latter choice for the 
fixed-order renormalization scale is motivated by the fact that 
the relative momentum of the top pair is the scale governing the
Coulomb singularities contained in the fixed-order expansion close to the
large Higgs energy endpoint. This choice for the fixed-order
renormalization scale is therefore the more appropriate one near the
Higgs energy endpoint. The results in Figs.~\ref{fig3} demonstrate 
the overlap between the ${\cal O}(\alpha_s)$ fixed-order
prediction and the NLL nonrelativistic one in the region where the
$t\bar t$ relative velocity is approximately 0.2. (The Higgs energy
with $v=0.2$ is indicated in each panel by the solid vertical line.) 
The overlap improves for increasing c.m.\,energies or decreasing Higgs
masses. This indicates that in the overlap regions the higher order
contributions summed in the nonrelativistic prediction and the higher
order relativistic corrections contained in the fixed-order result are
both small. For smaller c.m.\,energies or increasing Higgs 
masses, on the other hand, the NLL nonrelativistic predictions tend to
lie  slightly above the ${\cal O}(\alpha_s)$ fixed-order results (for 
$\mu=\sqrt{s}v$) illustrating the impact of the higher order
corrections to each type of expansion. The discrepancy, however,
remains comparable to the uncertainties estimated from the
renormalization parameter variation of the NLL nonrelativistic
prediction. We therefore conclude that the renormalization parameter
variation of the NLL order nonrelativistic prediction should provide a
realistic estimate of the theoretical uncertainties in the large Higgs
energy region. The results in Figs.\,\ref{fig3} also demonstrate that
the region of parameter space where the top quark pair is
nonrelativistic increases for smaller c.m.\,energies (or larger Higgs
masses). 

Let us now discuss the numerical impact of the nonrelativistic 
contributions in the large Higgs energy region on the total cross
section. In Tab.\,\ref{tab3} the importance of the summation of the
Coulomb singularities and the logarithms of the top quark velocity is
analyzed numerically for various choices of the c.m.\,energy and the 
Higgs mass. For all cases the top quark mass $m_t^{\rm 1S}=180$~GeV
is used and the other parameters are fixed as in Figs.~\ref{fig3}.
See the caption for details on the various entries. 
The numbers for $\sigma(\mbox{NLL})$, which are determined from
combining the NLL nonrelativistic predictions in the Higgs energy end
point for $|v|<0.2$ with the fixed-order ${\cal O}(\alpha_s)$
prediction for smaller Higgs energies with $|v|>0.2$, represent the
currently  most complete predictions for the total cross section of
the process  $e^+e^-\to t\bar t H$ as far as QCD corrections are
concerned. For $\sigma(\mbox{NLL})$ we have also given our estimate
for the theoretical error. 
For the fixed-order contribution
($v>0.2$) we have estimated the uncertainty by taking the maximum of
the shifts obtained from varying 
$\mu$ in the ranges $[\sqrt{s},2 \sqrt{s}]$, $[\sqrt{s},\sqrt{s}/2]$ 
and $[\sqrt{s}v,\sqrt{s}]$; for the nonrelativistic contribution in
the end point we have assumed an uncertainty of 5\% for all cases. 
For the numbers displayed in Tab.~\ref{tab3} both uncertainties were added 
linearly.

The results show that the enhancement of the cross section due to the
summations in the large Higgs  
energy region is particularly important for smaller c.m.\,energies 
and larger Higgs masses, when the portion of the phase space where the 
nonrelativistic expansion has to be applied is large. Here, the higher
order summations contained in the nonrelativistic expansion can be
comparable to the already sizeable ${\cal O}(\alpha_s)$ fixed-order
corrections and enhance the cross section further. This is advantageous
for top Yukawa coupling measurements for the lower c.m.\,energies 
accessible in the first phase of the ILC experiment. 
For higher c.m.\,energies the effect of the nonrelativistic summations
of contributions from beyond ${\cal O}(\alpha_s)$ is less pronounced
and decreases to the one-percent level for c.m.\,energies above
$700$~GeV. For all cases, except for very 
large c.m.\,energies around $1000$~GeV, however, the shift caused by
the terms that are summed up in the
nonrelativistic expansion exceeds the theoretical
error~\cite{farrellhoang1}. 

In Table~\ref{tab3}, in the last column, the ratio of the NLL
nonrelativistic cross section and the 
${\cal O}(\alpha_s)$ fixed-order cross section (with the approximation
$m_t^{\rm pole} = m_t^{\rm 1S}$) for $|v|<0.2$ is also
shown. Interestingly, the higher order summations lead 
to correction factors ranging between about $1.7$ and $1.8$ that are only
very weakly dependent of the c.m.\,energy and the Higgs mass. 
This fact might prove useful for rough
implementations of nonrelativistic $t\bar t$ effects in other high energy 
processes.

\appendix


\end{document}